\documentclass[twocolumn,showpacs,amsmath,amssymb,prb]{revtex4} 

\usepackage{graphicx}
\usepackage{epsf}

\begin{document}

\title{Gapped Heisenberg spin chains in a field}

\author{Rob Hagemans$^1$, Jean-S\'ebastien Caux$^1$ \footnote{Author to whom correspondence should be addressed.} and Ute L{\"o}w$^2$}

\affiliation{
$^1$ITFA, U. of Amsterdam, Valckenierstraat 65, 1018 XE Amsterdam, The Netherlands \\
$^2$ Institut f{\"u}r Theoretische Physik, U. zu K{\"o}ln, Z{\"u}lpicher Strasse 77, 50937 K{\"o}ln, Germany 
}

\date{\today}

\begin{abstract}
We consider the fully anisotropic Heisenberg spin-$1/2$ antiferromagnet in a uniform 
magnetic field, whose ground-state is characterized by broken spin rotation symmetry and gapped spinon 
excitations.  We expand on a recent mean-field approach to the problem by incorporating 
fluctuations in a loop expansion.  Quantitative results for the magnetization, 
excitation gap and specific heat are obtained.  We compare our predictions with new DMRG and
exact diagonalization data and,
for zero field, with the exact solution of the $\mbox{XYZ}$ spin chain from the Bethe Ansatz.
\end{abstract}


\maketitle

\section{Introduction}
For many years now, low-dimensional quantum systems have been a hotbed of theoretical and
experimental novelties.  Quasi one-dimensional spin systems have by now become
fertile laboratories for the investigation of many interesting phenomena stemming 
from strong correlations, an important example being charge fractionalization of physical excitations.
Overall, spin-$1/2$ Heisenberg chains count among the most important paradigms.  This 
stems first from the existence of many exact results on their ground-state, excitations and 
thermodynamics in a variety of cases, but also on the growing number of experimental
situations in which the effective theory boils down to a one-dimensional exchange
Hamiltonian.  

Exact results on quantum spin chains (by which we mean solvability at the lattice level)
are constrained to either 
easy-plane anisotropy in nonzero longitudinal field (so spin projection along the field axis is a good quantum
number, and spinon excitations are well-defined) (for a full exposition of this well-known subject, 
see {\it e.g.} [\onlinecite{KorepinBOOK,TakahashiBOOK}]), 
or zero field and general anisotropy \cite{BaxterAP70}.
Recently, however, there has been renewed interest in the study of spin chains in the more
general case of transverse field and anisotropy [\onlinecite{DmitrievPRB65,CauxPRB68,Dmitriev0403035,CapraroEPJB29}],
where all symmetries are broken and exact solutions are out of reach.
Most notably, the main motivation for the present study originates from
extensive work done on the field-dependent ground-state properties and excitation dynamics of
quasi-one-dimensional anisotropic antiferromagnets like
$\mbox{Cs}_2\mbox{CoCl}_4$ \cite{KenzelmannPRB65}.  

We will consider 
the Hamiltonian of the generally anisotropic Heisenberg chain in a field given by
\begin{eqnarray}
H \!=\! \sum_{j = 1}^N \left[ J_x S_j^x S_{j+1}^x \!+\! J_y S_j^y S_{j+1}^y 
\!+\! J_z S_j^z S_{j+1}^z \!-\! H_z S^z_j \right]
\label{Hspins}
\end{eqnarray}
where the spin-1/2 operators $S_j^{\alpha}$ obey the commutation
relations $\left[ S_j^{\alpha}, S_{j'}^{\beta} \right] = i \delta_{j j'} \epsilon^{\alpha \beta \gamma} 
S^{\gamma}_j$.  
This model shares many features with somewhat simpler and better understood exchange Hamiltonians
like the $\mbox{XY}$ \cite{LiebAP16} or $\mbox{XXZ}$ \cite{KorepinBOOK,TakahashiBOOK} models.  As far as the $\mbox{XY}$
model is concerned, many exact results are known in the presence of a field \cite{NiemeijerP36},
like the existence of an Ising-class disordering transition at a critical value of the field.
General space- and time-dependent correlation functions can be obtained by mapping to gapped 
noninteracting fermions
\cite{NiemeijerP36,McCoyPRA4,JohnsonPRA4,VaidyaPA92,TaylorPRB28}.

On the other hand, it is well-known that the isotropic antiferromagnetic Heisenberg chain, and more generally
the $\mbox{XXZ}$ ($J_x = J_y \neq J_z$) chain in a field parallel to the anisotropy (with $-1 < J_z/J_x \leq 1$), is a gapless
system with no long-range order and an unbroken $U(1)$ symmetry.  
Once again, many results are known on the thermodynamics \cite{KorepinBOOK,TakahashiBOOK} while
recent progress has been achieved on correlation
functions both on the lattice \cite{KitanineNPB641} and in the scaling limit \cite{LukyanovNPB654}.

For the fully anisotropic model in a field (\ref{Hspins}), however,
precious little is known.  In zero field, the model becomes exactly solvable \cite{BaxterAP70}, meaning that
thermodynamic properties are computable \cite{TakahashiBOOK}.  Correlation functions in zero field
are still unknown in general (with the exception of the Baxter-Kelland formula \cite{BaxterJPC7,JimboJPA26} for
the one-point function giving the staggered magnetization), although some special cases have been studied
\cite{RoldanPA136,JohnsonPRA8}.  
The distinguishing feature of (\ref{Hspins}) is that
one can find two independent origins for a gap:  first, in zero field, the very presence
of structural anisotropy breaks the symmetry right from the start.
One could also break the symmetry by applying a field
transverse to the anisotropy (like in the case of an $\mbox{XXZ}$ magnet with field along the
$y$ direction).  In both
of these cases, long-range N{\'e}el order develops.  However, the noncommuting
nature of the field with respect to the anisotropy and the fact that the field strength is 
tunable lead to the existence of quantum critical
behaviour in the system:  for large enough fields, fluctuations 
become strong enough to disorder the ground-state, leading to an Ising-class critical 
point at a field $H = H_c$ (whose value depends on the anisotropy).  This holds true
for a general anisotropic chain, studied in [\onlinecite{KurmannPA112}], in which 
bounds were obtained for the critical field, and rigorous results 
were derived on classical ground states for particular values
of the field (meaning that the ground-state becomes a two-sublattice N{\'e}el-type
state, with nonzero long-range antiferromagnetic order).

Exact results being inaccessible for now, one must rely on alternative methods to
compute quantities of interest like the magnetization along the field, transverse order and
objects like the dynamical structure factor.  
Close to a gapless, solvable point, 
one can rely on approaches like bosonization to obtain various scaling results
({\it e.g.} weak field and/or small anisotropy limits studied in \onlinecite{CauxPRB68}).  
However, the presence of a gap justifies 
the hope that a mean-field scheme might work out relatively well as a quantitative theory
provided the correct starting point is identified.
This was initiated in [\onlinecite{DmitrievPRB65}] for the $\mbox{XXZ}$ chain in a transverse field,
where the phase diagram and various scaling results were obtained.  Correlation functions were 
afterwards studied in [\onlinecite{CauxPRB68}] using the equivalence to the $\mbox{XY}$ model in a field,
and through the use of perturbative RG shemes based on sine-Gordon form factors.  
Some of these results are in fact immediately generalizable to the $\mbox{XYZ}$ chain in a field
that we purport to study.  One thing that was missing, however, was an
estimate of the strength of fluctuations, which is of particular importance 
in view of potential quantitative comparisons with experiments.  For low fields,
the mean-field approach failed by construction to properly describe the restoration of the $U(1)$
symmetry of the $\mbox{XXZ}$ chain.  The mean-field staggered magnetization, for example, failed to
vanish at zero field, in clear contradiction with the exact result.  
The field dependence of the gap also failed for small fields, disagreeing with both experiments
and numerical results using the Density Matrix Renormalization Group (DMRG).  
One might ask whether DMRG or other quasi-exact methods would suffice to describe all aspects of experiments, but its main
drawback is that it is only very useful as a tool for investigating 
properties of the ground-state and lowest-lying states, and is not yet an
economical tool for dynamical quantities like spin correlation functions, measurable using
neutron scattering.  Our general idea is 
therefore quite simple:  to carefully compute fluctuations around mean-field to two loops, 
providing an improved setup from which values for the thermodynamic (and eventually dynamic) properties of the system, 
and quantitative estimates of the precision of the mean-field theory can be obtained 
through the use of intensive numerics on the resulting equations.  

This paper is organized as follows.  In section II, the basic mean-field setup is presented.
Fluctuations around mean-field are computed in section III, and used to determine corrections to
physical parameters, the specific heat, and the excitation gap.  
Section IV presents a comparison of our approximations in zero field with the
exact results available from the Bethe Ansatz, and conclusions are collected in Section V.

\section{Setup}
Written as it is, (\ref{Hspins}) is not tractable in any simple way.  To make progress, we
move from a spin basis to a simpler fermion one 
by using the Jordan-Wigner transformation $S^z_j = c^{\dagger}_j c_j - 1/2$, $S^{\dagger}_j = 
e^{i \pi \sum_{k = 1}^{j-1} c^{\dagger}_k c_k} c^{\dagger}_j$, under which the Hamiltonian
gets transformed to a theory of interacting fermions in the presence of a superconducting-like
Cooper pairing term:
\begin{eqnarray}
H \!=\! \sum_{j=1}^N \frac{J_+}{2} (c^{\dagger}_j c_{j+1} \!+\! c^{\dagger}_{j+1} c_j)
+ \frac{J_-}{2} (c^{\dagger}_{j} c^{\dagger}_{j+1} \!+\! c_{j+1} c_j)
+ \nonumber \\
+ J_z (c^{\dagger}_j c_j - 1/2)(c^{\dagger}_{j+1} c_{j+1} - 1/2) - H_z (c^{\dagger}_j c_j - 1/2) \hspace{0.2cm}
\label{Hfermions}
\end{eqnarray}
where $J_{\pm} = (J_x \pm J_y)/2$.  
We impose periodic boundary conditions along the lattice on the spin operators, so 
$S^{\alpha}_{j+N} = S^{\alpha}_j$.  The fermion operators are thus taken to have antiperiodic
boundary conditions.  We shall restrict the exchange parameters to
the domain $J_x \geq J_z \geq 0, J_x \geq |J_y| \geq 0$, therefore concentrating on the antiferromagnetic regime.  
Note that (\ref{Hspins}) is not quite the most general anisotropic exchange
Hamiltonian in a field that we could have started from:  pointing the (field, or equivalently the ) anisotropy in 
a completely general
direction would entail terms of the form $S_j^{\alpha} S_{j+1}^{\beta}$ with $\alpha \neq \beta$.
The absence of $x,z$ and $y,z$ exchange terms is here required to ensure a local fermionic theory.

The interacting fermion Hamiltonian (\ref{Hfermions}) is not exactly solvable because of the 
simultaneous presence of the pairing-like term $J_-$, originating from the transversality of the magnetic field
and $x,y$ anisotropy, and the external magnetic field.  We therefore  
use the approximation scheme proposed in [\onlinecite{DmitrievPRB65}] and elaborated
in [\onlinecite{CauxPRB68}].  Let us recall here the basics of this setup.
First, the fermionic interaction term is decomposed via the (site independent) mean-field parameters 
\begin{eqnarray}
{\cal M}_0 \!=\! \langle (c^{\dagger}_j c_j - 1/2) \rangle_0, \hspace{0.2cm}
{\cal K}_0 \!=\! \langle c^{\dagger}_{j+1} c_j \rangle_0, \hspace{0.2cm}
{\cal P}_0 \!=\! \langle c_{j+1} c_j \rangle_0
\end{eqnarray}
corresponding to the magnetization, kinetic and pairing terms respectively.  
The angular brackets with subscript $0$ signify thermal averages using the mean-field states.
The mean-field Hamiltonian becomes an exactly solvable free fermion theory (corresponding to an
effective anisotropic $\mbox{XY}$ model in a field), 
\begin{eqnarray}
H_{MF} \!=\! \sum_{j=1}^N \!\frac{t}{2} (c^{\dagger}_j c_{j+1} \!+\! c^{\dagger}_{j+1} c_j)
\!+\! \frac{\Delta}{2} (c^{\dagger}_{j} c^{\dagger}_{j+1} \!+\! c_{j+1} c_j) \!-\! \nonumber \\
- h (c^{\dagger}_j c_j - 1/2) 
+ \bar{e}_0
\hspace{2cm}
\label{HMF}
\end{eqnarray}
in which $\bar{e}_0 = J_z (-{\cal M}_0^2 + {\cal K}_0^2 - {\cal P}_0^2)$, and  
\begin{eqnarray}
t = J_+ \!-\! 2J_z {\cal K}_0, \hspace{0.2cm} \Delta = J_- \!+\! 2J_z {\cal P}_0, 
\hspace{0.2cm} h = H_z \!-\! 2J_z {\cal M}_0.
\end{eqnarray}
The parameters $t, \Delta$ and $h$ are thus simply interpreted physically as effective
bandwidth, effective superconductivity-like pairing, and effective chemical potential.  The
ground-state is therefore $BCS$-like and contains all particle numbers.  
Performing a Fourier transform to momentum space using $c_j = \frac{1}{\sqrt{N}} \sum_{n = 1}^N e^{i k_n j} c_{k_n}$, 
$k_n = \frac{2\pi}{N}(n - 1/2)$, $c_{j+N} = - c_j$
and choosing $N$ even for convenience, we can write
\begin{eqnarray}
\!\!H_{MF} \!=\! \sum_{n =1}^{N/2} 
{\bf c}^{\dagger}_{k_n} \!\!
\left( \begin{array}{cc} t \cos k_n - h & i \Delta \sin k_n \\
-i \Delta \sin k_n & -t\cos k_n + h \end{array} \right) \!
{\bf c}_{k_n} + \bar{e}_0
\end{eqnarray}
using the Nambu spinor 
${\bf c}^{\dagger}_{k_n} = \left( \begin{array}{cc} c^{\dagger}_{k_n} & c_{-k_n} \end{array} \right)$.
The theory can easily be rediagonalized as
\begin{eqnarray}
H_{MF} = \sum_{n=1}^{N/2} {{\bf c}'}^{\dagger}_{k_n} \sigma^z \omega(k_n) {\bf c}'_{k_n} + \bar{e}_0
\end{eqnarray}
with Bogoliubov-de Gennes transformation parameters
\begin{eqnarray}
{\bf c}'_{k} = {\bf B}_{k} {\bf c}_k, \hspace{0.3cm} 
{\bf B}_{k} = \left( \begin{array}{cc} \alpha_+(k) & i \alpha_-(k) \\
i\alpha_-(k) & \alpha_+(k) \end{array} \right),
\end{eqnarray}
\begin{eqnarray}
\alpha_{\pm} (k) = \frac{1}{\sqrt{2}} \left[ 1 \pm \frac{t \cos k - h}{\omega(k)}\right]^{1/2},
\end{eqnarray}
and $\alpha_{\pm} (-k) = \pm \alpha_{\pm} (k)$.  The dispersion relation is
\begin{eqnarray}
\omega(k) = \left[ (t\cos k - h)^2 + \Delta^2 \sin^2 k \right]^{1/2}
\label{omegak}
\end{eqnarray}
and makes manifest the presence of the gap in the theory.  Now, in terms of the new fermions,
we sit always precisely at half-filling, and the ground-state is characterized by a fully
filled lower band.  Particle-hole excitations correspond to spinon-like excitations in the original
spin model:  exactly as in the $\mbox{XY}$ model, due to energy-momentum conservation 
two-particle processes dominate below the critical field, whereas above it one-particle
processes give the strongest contribution to correlation functions. 

The thermodynamic properties of the model at the mean-field level are easily 
obtained from the mean-field free energy
per site at inverse temperature $\beta$,
\begin{eqnarray}
f_0 = \frac{-1}{N\beta} \ln Z_{MF}, \hspace{0.3cm} Z_{MF} = \mbox{Tr} e^{-\beta H_{MF}}.
\end{eqnarray}
Explicitly, in the limit $N \rightarrow \infty$, we get
\begin{eqnarray}
f_0 = \bar{e}_0 + \frac{-2}{\beta} \int_0^{\pi} \frac{dk}{2\pi} \ln \cosh \frac{\beta \omega(k)}{2}.
\end{eqnarray}
The mean-field setup is completed by requiring that the following 
self-consistency equations be satisfied:
\begin{eqnarray}
&&\!\!\!\!\!\!{\cal M}_0 \!+\! 1/2 \!=\! \int_0^{\pi} \!\!\frac{dk}{2\pi} \left[1 \!-\! (1 \!-\! 2n_{\beta} (\omega(k)))
(\alpha_+^2(k) \!-\! \alpha_-^2(k)) \right], \nonumber \\
&&\!\!\!\!\!\!{\cal K}_0 \!=\! \int_0^{\pi} \!\!\frac{dk}{2\pi} \cos k \left[1 \!-\! (1 \!-\! 2n_{\beta} (\omega(k)))
(\alpha_+^2(k) \!-\! \alpha_-^2(k)) \right], \nonumber \\
&&\!\!\!\!\!\!{\cal P}_0 \!=\! -\!\int_0^{\pi} \!\!\frac{dk}{2\pi} \sin k (1 \!-\! 2n_{\beta} (\omega(k)))
\alpha_+ (k) \alpha_- (k)
\end{eqnarray}
with $n_{\beta} (a) = (1 + e^{\beta a})^{-1}$.  These are easily solved numerically with a simple iterative
procedure.  One important consequence is that all mean-field parameters, and therefore all effective potentials
appearing in the mean-field Hamiltonian, acquire a temperature dependence.  
The average energy per site in the mean-field approximation is itself simply given by
\begin{eqnarray}
e_0 \!\!={\bar e}_0\!\! -\!\!\int_0^{\pi} \!\!\frac{dk}{2\pi} \omega (k) \tanh \frac{\beta \omega(k)}{2} 
\nonumber \\
= t {\cal K}_0 \!+\! \Delta {\cal P}_0 \!-\! h {\cal M}_0 + {\bar e}_0.
\end{eqnarray}
The mean-field specific heat is defined for fixed external field $H_z$, 
$c  = \frac{\partial {e}_0}{\partial T} |_{H_z}$.  

\section{Fluctuations around mean-field}
The fluctuations around mean-field can be described using standard perturbative expansions by reinstating 
the original fermionic interaction term
in the Hamiltonian (\ref{HMF}).  In detail, the perturbation that we have to deal with reads
\begin{eqnarray}
H_{FL} = \frac{J_z}{N} \!\!\!\!\sum_{n_1, n_2, n_3 = 1}^N \cos k_{n_3}^0
:\!c^{\dagger}_{k_{n_1 + n_3}} c_{k_{n_1}} c^{\dagger}_{k_{n_2 - n_3}} c_{k_{n_2}}\!\!:
\end{eqnarray}
in which the symbol $::$ signifies that self-contractions are absent in perturbation theory, 
and where we introduce the notation $k_n^0 = \frac{2\pi}{N} n$.  We are being pedantic with 
the momenta, since we keep all finite $N$ corrections in the computations (this turns out to help
the numerics perceptibly, since we can realistically plot our functions for $N$ only of the order of
a few hundred).  

Since we are interested in equilibrium thermodynamic quantities, 
the perturbative computations are most easily tackled using the 
imaginary time ($\tau = it$) formalism.  First, we define the Matsubara Green's functions
as appropriate averages over the mean-field theory (represented everywhere by $\langle ... 
\rangle_0$).  We need two types of functions, one representing normal correlations and
the other, pairing correlations:  
\begin{eqnarray}
{\cal G}_0 (k_n, -i\tau) \delta_{k_n, k_{n'}} = - \langle \mbox{T}_{\tau} \left\{ c_{k_n} (-i\tau) 
c^{\dagger}_{k_{n'}} (0) \right\} \rangle_0 \nonumber \\
{\cal F}_0 (k_n, -i\tau) \delta_{k_n + k_{n'}, 0} = \langle \mbox{T}_{\tau} \left\{ c_{k_n} (-i\tau) 
c_{k_{n'}} (0) \right\} \rangle_0.
\end{eqnarray}
The most convenient representation, exhibiting the poles in the normal and Cooper channels, is obtained after 
Fourier transforming to frequency space (we use the definitions 
$f(i \omega_{\alpha}) = \int_0^{\beta} d\tau e^{i \omega_{\alpha} \tau} f(-i\tau)$,
$\omega_{\alpha} = \frac{2\pi}{\beta} (\alpha + 1/2)$, $\alpha \in \mathbb{Z}$)
\begin{eqnarray}
&&{\cal G}_0 (k, i\omega_{\alpha}) = \frac{\alpha_+^2 (k)}{i\omega_{\alpha} - \omega(k)} 
  + \frac{\alpha_-^2 (k)}{i\omega_{\alpha} + \omega(k)},
\nonumber \\
&&{\cal F}_0 (k, i\omega_{\alpha}) = 2 \alpha_+ (k) \alpha_- (k) \frac{i\omega (k)}{\omega_{\alpha}^2 + \omega^2 (k)}.
\end{eqnarray}
All physical quantities of interest can be computed using these functions, in particular the 
unperturbed mean-field parameters:
\begin{eqnarray}
&&{\cal M}_0 = \frac{1}{\beta N} \sum_{n=1}^N \sum_{\alpha \in \mathbb{Z}} {\cal G}_0 (k_n, i\omega_{\alpha}), 
\nonumber \\
&&{\cal K}_0 = \frac{1}{\beta N} \sum_{n=1}^N \sum_{\alpha \in \mathbb{Z}} \cos k_n {\cal G}_0 (k_n, i\omega_{\alpha}), 
\nonumber \\
&&{\cal P}_0 = \frac{i}{\beta N} \sum_{n=1}^N \sum_{\alpha \in \mathbb{Z}} \sin k_n {\cal F}_0 (k_n, i\omega_{\alpha}).
\label{MFparamsGF}
\end{eqnarray}

The full Green's functions in imaginary time are defined as
\begin{eqnarray}
&&{\cal G} (k, -i\tau) = - \langle \mbox{T}_{\tau} \left\{ c_k (-i\tau) c_k^{\dagger} (0) 
\right\} \rangle, \nonumber \\
&&{\cal F} (k, -i\tau) = \langle \mbox{T}_{\tau} \left\{ c_k (-i\tau) c_k (0) 
\right\} \rangle
\end{eqnarray}
where the angular brackets denote a thermal average over the full Hamiltonian $H = H_{MF} + H_{FL}$.
The perturbation expansion can then be formally written down as the series
(here in the normal Green's function)
\begin{eqnarray}
&&{\cal G} (k, -i\tau) = - \sum_{n=0}^{\infty} \frac{(-1)^n}{n!} \int_0^{\beta} 
d\tau_1 ... d\tau_n \times \hspace{2cm}\nonumber \\ 
&&\times \langle \mbox{T}_{\tau}
\left\{ c_k (-i\tau) H_{FL} (-i\tau_1) ... H_{FL} (-i\tau_n) c^{\dagger}_k (0) \right\} \rangle_0
\end{eqnarray}
in which all correlators are now evaluated at mean-field level.
To first order in perturbation theory, fluctuations do not correct the mean-field Green's
functions (by definition).  
The first corrections appear at second order, where for each type of Green's function
12 different diagrams can be written down.  Summing these after chopping external legs 
provides a computation of the self-energies to two loops.
After a relatively lengthy computation, the self-energies
to second order (as identified by the superscript) can be written down as
(after summation over internal Matsubara frequencies, only the sum over the two internal
momenta $k_1, k_2$ remains) 
\begin{eqnarray}
\Sigma_N^{(2)} (k, i\omega) &=& -\frac{4 J_z^2}{N^2} \sum_{k_1, k_2} T_N (k, i\omega | k_1, k_2), \nonumber \\
\Sigma_S^{(2)} (k, i\omega) &=& i\frac{4 J_z^2}{N^2} \sum_{k_1, k_2} T_S (k, i\omega | k_1, k_2),
\label{Sigma2}
\end{eqnarray}
where the ``normal'' and ``Cooper'' self-energies both receive contributions from normal and Cooper
channels.  Explicitly, we write those contributions as
\begin{widetext}
\begin{eqnarray}
\!\!\!\!T_N (k, i\omega | k_1, k_2) \!\!&=&\!\!\!\! \sum_{\sigma_1 \sigma_2 \sigma_3} (1-\tilde{\delta})
\left[ C^{NNN}_{\sigma_1 \sigma_2 \sigma_3} 
(k | k_1, k_2) f_1 (k | k_1, k_2) + C^{SSN}_{\sigma_1 \sigma_2 \sigma_3} f_2 (k | k_1, k_2) \right]
D_{\sigma_1 \sigma_2 \sigma_3} (k, i\omega | k_1, k_2) + \tilde{\delta} T_N^c, \nonumber \\
\!\!\!\!T_S (k, i\omega | k_1, k_2) \!\!&=&\!\!\!\! \sum_{\sigma_1 \sigma_2 \sigma_3} (1 - \tilde{\delta}) 
\left[ C^{SSS}_{\sigma_1 \sigma_2 \sigma_3} 
(k | k_1, k_2) f_1 (k | k_1, k_2) + C^{NNS}_{\sigma_1 \sigma_2 \sigma_3} f_2 (k | k_1, k_2) \right]
D_{\sigma_1 \sigma_2 \sigma_3} (k, i\omega | k_1, k_2) + \tilde{\delta} T_S^c,
\label{T}
\end{eqnarray}
and the external frequency $i\omega$-dependent part is 
\begin{eqnarray}
D_{\sigma_1 \sigma_2 \sigma_3} (k_n, i\omega | k_{n_1}, k_{n_2}) = \frac{[n_{\beta} (\sigma_3 \omega(k_{n_2})) 
- n_{\beta} (\sigma_2 \omega(k_{n_1 + n_2})] [n_{\beta}^B (\sigma_2 \omega(k_{n_1 + n_2}) - \sigma_3 \omega(k_{n_2}))
+ n_{\beta} (\sigma_1 \omega(k_{n_1 + n}))]}{i\omega - \sigma_1 \omega(k_{n_1 + n}) + \sigma_2 \omega(k_{n_1 + n_2})
- \sigma_3 \omega(k_{n_2})}
\end{eqnarray}
\end{widetext}
where $n_{\beta}^B(a) = (e^{\beta a} - 1)^{-1}$.  The terms $T_{N,S}^c$ are finite $N$ corrections which
are detailed below.  These have been explicitly kept in the numerics, where they noticeably accelerate 
convergence towards the thermodynamic limit.  The parameter
\begin{equation}
\tilde{\delta} = \delta_{\sigma_2 \sigma_3} \delta_{\omega(k_{n_1 + n_2}), \omega(k_{n_2})}
\end{equation}
comes from the double-pole contributions in the internal Matsubara frequency summation.  It
can of course be put to zero in the strict $N \rightarrow \infty$ limit.  The momentum-dependent parameters in 
(\ref{T}) are given by various combinations of the Bogoliubov-de Gennes transformation parameters, namely
\begin{eqnarray}
&&C^{NNN}_{\sigma_1 \sigma_2 \sigma_3} = \alpha_{\sigma_1}^2 (k_{n_1 + n}) 
\alpha_{\sigma_2}^2 (k_{n_1 + n_2}) \alpha_{\sigma_3}^2 (k_{n_2}), \nonumber \\
&&C^{SSN}_{\sigma_1 \sigma_2 \sigma_3} = \sigma_1 \sigma_2 \alpha_{+-} (k_{n_1 + n}) 
\alpha_{+-} (k_{n_1 + n_2}) \alpha_{\sigma_3}^2 (k_{n_2}), \nonumber \\
&&C^{SSS}_{\sigma_1 \sigma_2 \sigma_3} = \sigma_1 \sigma_2 \sigma_3 \alpha_{+-} (k_{n_1 + n}) 
\alpha_{+-} (k_{n_1 + n_2}) \alpha_{+-} (k_{n_2}), \nonumber \\
&&C^{NNS}_{\sigma_1 \sigma_2 \sigma_3} = \sigma_3 \alpha_{\sigma_1}^2 (k_{n_1 + n}) 
\alpha_{\sigma_2}^2 (k_{n_1 + n_2}) \alpha_{+-} (k_{n_2})
\end{eqnarray}
where $\alpha_{+-} (k) = \alpha_+ (k) \alpha_-(k)$, and we have suppressed the arguments $(k | k_1, k_2)$ 
for clarity.  The momentum weight factors are trigonometric functions given by (we remind the reader that $k_n^0 = \frac{2\pi}{N}n$;
moreover, since we sum over the full Brillouin zone, the Umklapp terms are explicitly included in our expressions)
\begin{eqnarray}
f_1 (k_n | k_{n_1}, k_{n_2}) = -\frac{(\cos k_{n_1}^0 - \cos k_{n_2 - n}^0)^2}{2}, \nonumber \\
f_2 (k_n | k_{n_1}, k_{n_2}) = (\cos k_{n_2 - n}^0 - \cos k_{n_1}^0) \times \nonumber \\
\times (\cos k_{n_2 - n}^0 \!-\! \cos k_{n_1 \!+\! n_2 \!+\! n \!+\! 1}^0). \hspace{0.5cm}
\end{eqnarray}
The finite $N$ correction terms included in the self-energy expressions (\ref{T}) are finally given by
\begin{eqnarray}
T_N^c &=& V^{NNN} + V^{SSN}, \nonumber \\
T_S^c &=& V^{SSS} + V^{NNS}
\end{eqnarray}
in which (for $A,B,C = N,S$) 
\begin{widetext}
\begin{eqnarray}
V^{ABC} = \left[ \frac{f^A_+ (k_1 + k)}{i\omega - \omega(k_1 + k)} + 
 \frac{f^A_- (k_1 + k)}{i\omega + \omega(k_1 + k)} \right] (f^B_+ (k_2) f^C_+ (k_2) 
+ f^B_- (k_2) f^C_- (k_2))
n_{\beta} (\omega(k_2) (1 - n_{\beta} (\omega(k_2)))
\end{eqnarray}
\end{widetext}
and where we have used the notations
\begin{eqnarray}
f^N_{\pm} (k) = \alpha_{\pm}^2 (k), \hspace{0.3cm} f^S_{\pm} (k) = \mp i \alpha_{+-} (k).
\end{eqnarray}

It is evident that the self-energies manifestly obey the same symmetries as the Green's functions, namely
\begin{eqnarray}
\Sigma_N (k, i\omega) &=& \Sigma_N (-k, i\omega), \nonumber \\
\Sigma_S (k, i\omega) &=& -\Sigma_S (-k, i\omega), \nonumber \\
\Sigma_S (k, -i\omega) &=& \Sigma_S (k, i\omega).
\end{eqnarray}

At this stage, a perturbative (partial) resummation can naturally and easily be performed using Dyson's equation.  We start
by defining the basic propagator matrix
\begin{eqnarray}
{\boldsymbol{\mathcal{G}}}_0 (k, i\omega) = \left( \begin{array}{cc} 
{\cal G}_0 (k, i\omega) & {\cal F}_0 (k, i\omega) \\
-{\cal F}_0 (k, i\omega) & -{\cal G}_0 (k, -i\omega) \end{array} \right).
\end{eqnarray}
In the same way, we can arrange the self-energies into the 
self-energy matrix
\begin{eqnarray}
{\boldsymbol{\Sigma}} (k, i\omega) = \left( \begin{array}{cc}
\Sigma_N (k, i\omega) & \Sigma_S (k, i\omega) \\
-\Sigma_S (k, i\omega) & - \Sigma_N (k, -i\omega) \end{array} \right).
\end{eqnarray}
The basic resummation is then performed by noting that the full propagator matrix fulfils the
Dyson equation
\begin{eqnarray}
{\boldsymbol{\cal G}} (k, i\omega) = {\boldsymbol{\cal G}}_0 (k, i\omega) + {\boldsymbol{\cal G}}_0 (k, i\omega) 
{\boldsymbol{\Sigma}} (k, i\omega) {\boldsymbol{\cal G}} (k, i\omega),
\end{eqnarray}
in turn leading to the full matrix propagator
\begin{eqnarray}
{\boldsymbol{\cal G}} (k, i\omega) = \left[  {\boldsymbol{\cal G}}_0^{-1} (k, i\omega) 
  - {\boldsymbol{\Sigma}} (k, i\omega) 
\right]^{-1}.
\end{eqnarray}
In terms of mean-field Green's functions and self-energies, this reads 
\begin{widetext}
\begin{eqnarray}
{\boldsymbol{\cal G}} (k, i\omega) = D^{-1} \left( \begin{array}{cc}
{\cal G}_0 (k, i\omega) + \mbox{Det}({\bf {\cal G}}_0 (k, i\omega)) \Sigma_N (k, -i\omega) &
{\cal F}_0 (k, i\omega) + \mbox{Det}({\bf {\cal G}}_0 (k, i\omega)) \Sigma_S (k, i\omega) \\
-{\cal F}_0 (k, i\omega) - \mbox{Det}({\bf {\cal G}}_0 (k, i\omega)) \Sigma_S (k, i\omega) &
-{\cal G}_0 (k, -i\omega) - \mbox{Det}({\bf {\cal G}}_0 (k, i\omega)) \Sigma_N (k, i\omega) 
\end{array} \right), 
\end{eqnarray}
\end{widetext}
with denominator
\begin{eqnarray}
D \!\!=\!\! 1 \!-\! \mbox{Tr} ({\boldsymbol{\cal G}}_0 (k,i\omega) {\boldsymbol{\Sigma}} (k, i\omega))
\!+\! \mbox{Det} ({\boldsymbol{\cal G}}_0 (k,i\omega) {\boldsymbol{\Sigma}} (k, i\omega)) \nonumber \\
\label{Dysondenom}
\end{eqnarray}
where the operators $\mbox{Tr}$ and $\mbox{Det}$ operate in the Nambu space, 
and where $\mbox{Det} ({\boldsymbol{\cal G}}_0 (k, i\omega)) = -[\omega^2 + \omega^2(k)]^{-1}$.
The two-loop approximation then corresponds to replacing the exact ${\boldsymbol{\Sigma}}(k, i\omega)$
by the two-loop result ${\boldsymbol{\Sigma}}^{(2)} (k, i\omega)$ (\ref{Sigma2}).

\subsection{Corrections to mean-field parameters}
As a first application of our formulas, 
the modifications to the mean-field parameters can be computed using our full Green's functions.
We define the general (off mean-field) averages 
\begin{eqnarray}
{\cal M} \!=\! \langle (c^{\dagger}_j c_j - 1/2) \rangle, \hspace{0.2cm}
{\cal K} \!=\! \langle c^{\dagger}_{j+1} c_j \rangle, \hspace{0.2cm}
{\cal P} \!=\! \langle c_{j+1} c_j \rangle,
\end{eqnarray}
which can (in complete analogy with the pure mean-field ones, see (\ref{MFparamsGF})) be obtained via the summation rules
\begin{eqnarray}
&&{\cal M} = \frac{1}{\beta N} \sum_{n=1}^N \sum_{\alpha \in \mathbb{Z}} {\cal G} (k_n, i\omega_{\alpha}), 
\nonumber \\
&&{\cal K} = \frac{1}{\beta N} \sum_{n=1}^N \sum_{\alpha \in \mathbb{Z}} \cos k_n {\cal G} (k_n, i\omega_{\alpha}), 
\nonumber \\
&&{\cal P} = \frac{i}{\beta N} \sum_{n=1}^N \sum_{\alpha \in \mathbb{Z}} \sin k_n {\cal F} (k_n, i\omega_{\alpha}).
\end{eqnarray}
Plots of these quantities as a function of field for various values of the anisotropy parameters are
given in figures (1-3).  In the case of the magnetization, it can be seen that the inclusion of fluctuations
within the framework has vastly improved the predictions of the theory all over the parameter space.  
For example, the relative error at $H_z = 1$ for $(J_x, J_y, J_z) = (1.0, 0.25, 1.0)$ has gone from 
12.5 percent for the mean-field versus DMRG result, to 1.9 percent for fluctuations versus DMRG.

\begin{figure}
\includegraphics[width=8cm]{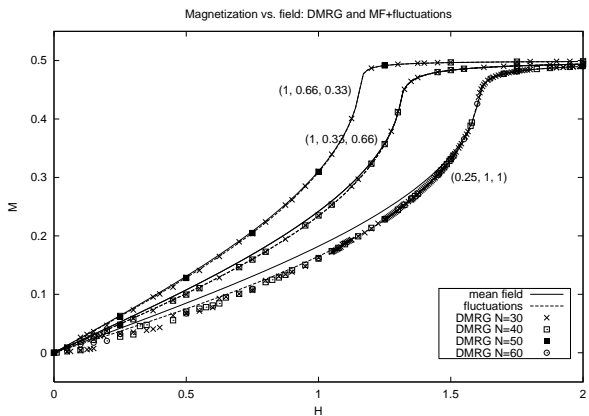}
\label{mag_H}
\caption{Magnetization as a function of field, for different anisotropy cases, 
computed using the basic mean-field setup, and with fluctuations.  The data points are
the DMRG results.}
\end{figure}
\begin{figure}
\includegraphics[width=8cm]{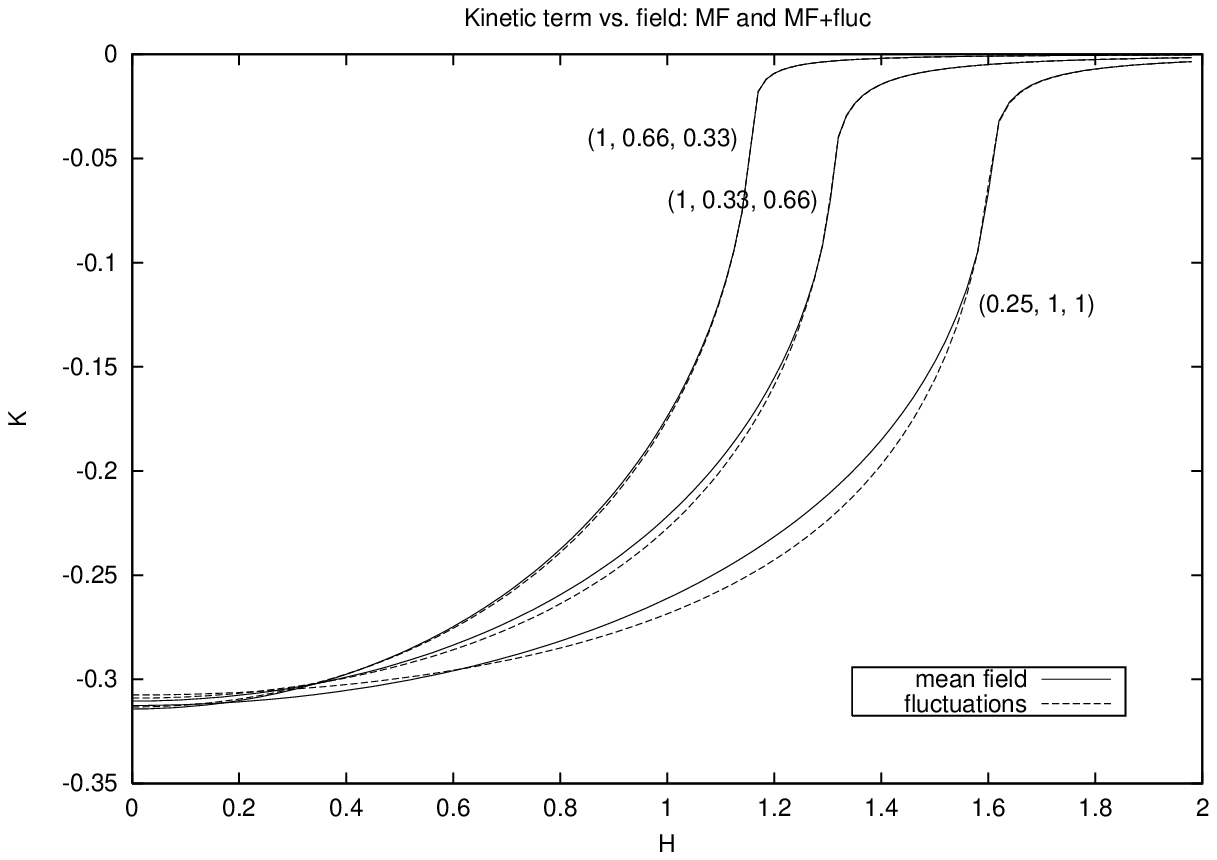}
\label{kin_H}
\caption{Kinetic term as a function of field, for different anisotropy cases, 
computed using the basic mean-field setup, and with fluctuations.}
\end{figure}
\begin{figure}
\includegraphics[width=8cm]{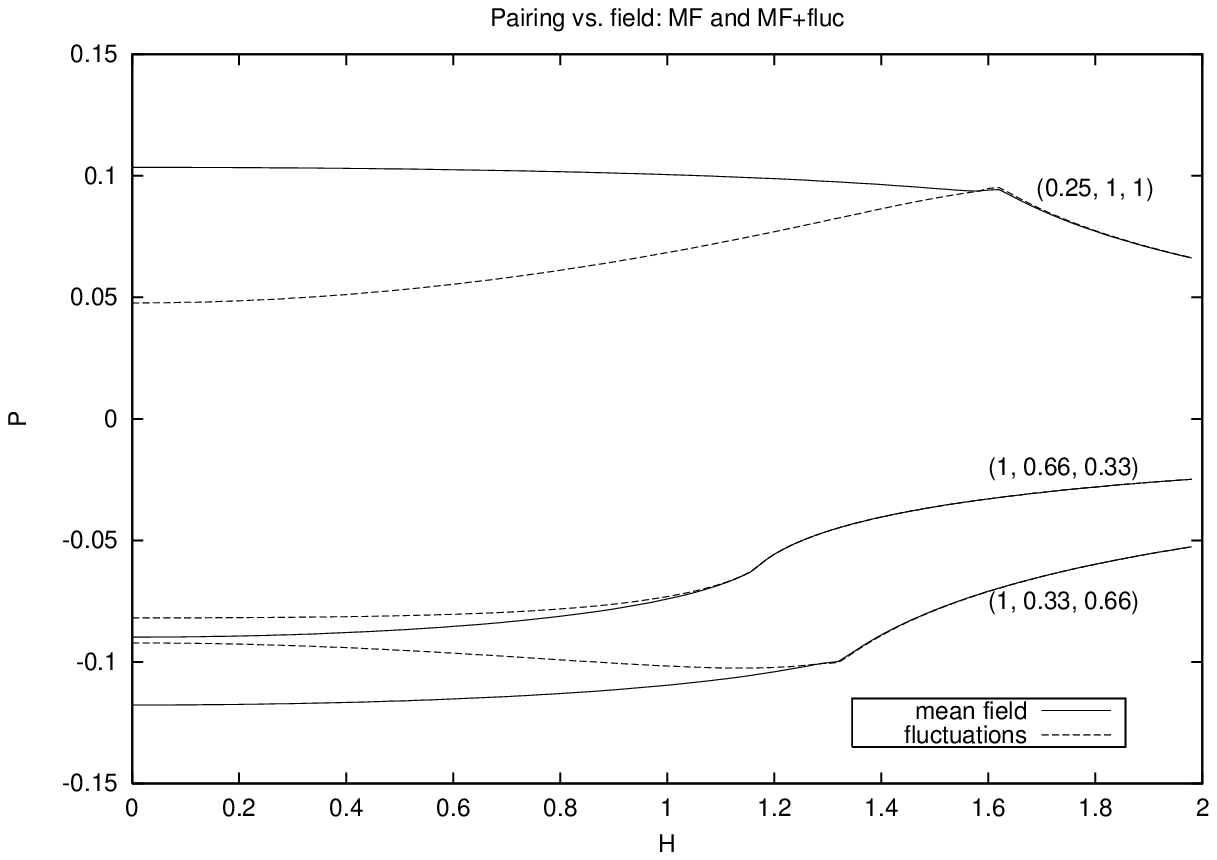}
\label{par_H}
\caption{Pairing term as a function of field, for different anisotropy cases, 
computed using the basic mean-field setup, and with fluctuations.}
\end{figure}

\subsection{Energy per site and specific heat}
\begin{figure}
\includegraphics[width=9cm]{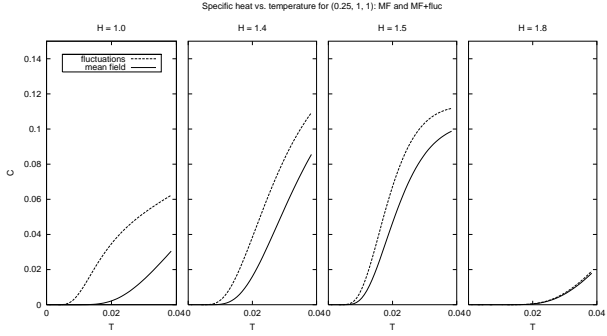}
\caption{Low-temperature specific heat for the $\mbox{XXZ}$ case with anisotropy of $0.25$, for different values of the external field.
Four plots are given, corresponding respectively to $H_z$ = $1.0, 1.4, 1.5$ and $1.8$.  In each graph, the mean-field and 
fluctuations-corrected values are given by the two separate curves.  The first three fields
are below the critical field, the fourth somewhat above.  Incorporating the fluctuations has reduced the gap.}
\label{cXXZ}
\end{figure}

\begin{figure}
\includegraphics[width=8cm]{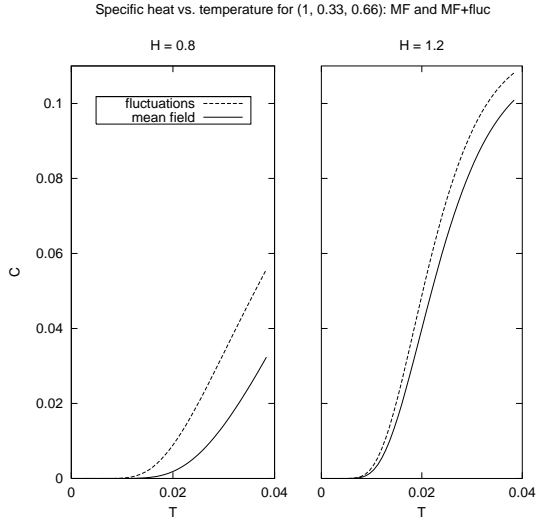}
\caption{Specific heat for the $\mbox{XYZ}$ case, for two values of the magnetic field (below the transition).  Both the mean-field
result and the fluctuations-corrected one are plotted for small temperatures.}
\label{c2}
\end{figure}

\begin{figure}
\includegraphics[width=8cm]{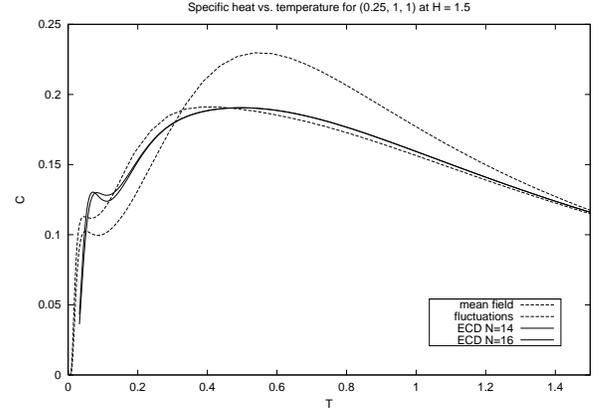}
\caption{Specific heat for the $\mbox{XXZ}$-like case, plotted over a larger temperature range.  Besides the
mean-field and fluctuations results, exact diagonalization ones are presented for lattice sizes $14$ and $16$.}
\label{c3}
\end{figure}

The mean energy per site can be simply obtained perturbatively either directly or through 
a linked cluster-like computation as
\begin{eqnarray}
e &=& e_0 + t \delta{\cal K} + \Delta \delta{\cal P} - h \delta{\cal M} + \nonumber \\
&&+ \frac{1}{4N\beta} \sum_{k,i\omega_n} \mbox{Tr} \left[ {\boldsymbol{\cal G}} (k,i\omega_n) 
{\boldsymbol{\Sigma}} (k, i\omega_n) \right]
\end{eqnarray}
in which $\delta {\cal K} = {\cal K} - {\cal K}_0$, etc.  
Substituting the two-loop results we have
derived for the matrix Green's function and self-energy, we can compute the mean energy
and more interestingly the specific heat as a function of field and temperature.
In figure (\ref{cXXZ}), the $\mbox{XXZ}$-like case with anisotropies $(1.0, 0.25, 1.0)$ is considered.
Since this system is gapless at zero field, this is the ``worst-case'' scenario considering that
our mean-field theory starts with a nonzero gap even at zero field.   
Both the straight mean-field, and the fluctuations-corrected values are plotted.  The gap computed
using fluctuations is lower than the mean-field one, as can be seen from the low-temperature onset
of the specific heat.  For external field $H_z = 1.8$, above the transition, the gaps are essentially
the same, and so are the specific heat curves.  In figure (\ref{c2}), the differences between the
mean-field result and the fluctuations-corrected one are milder, since the gap is bigger.  Once again,
the gap is reduced by the fluctuations.  Finally, figure (\ref{c3}) gives the specific heat through
a larger temperature domain for the $\mbox{XXZ}$-like case.  There is a rather large difference 
between the two theoretical curves (from mean-field and fluctuations) over the whole temperature range
plotted.  Comparing with the finite-size exact diagonalization (ECD) \cite{BF} results presented, it is clear that 
the mean-field result has been greatly improved by incorporating fluctuations, and that at low
temperatures ($T/J \leq 0.2$) the numerical curves tend to converge towards the theoretical ones 
when going to higher size (for higher temperatures, the finite-size results are essentially those of the continuum
limit).  

\subsection{Zero temperature excitation gap}
The mean-field prediction for the gap is easily found from the dispersion relation
(\ref{omegak}).  As mentioned previously, below the transition two-particle excitations yielding
incoherent spin-spin correlation functions dominate, and the gap is twice the single particle gap,
whereas above the transition the one-particle excitation sets the gap.  This yields
\begin{eqnarray}
\omega_{gap} = \left\{ \begin{array}{cc} 2 |\Delta| \sqrt{1 - \frac{h^2}{t^2 - \Delta^2}}, & |h| \leq |t - \Delta^2/t|, \\
2(|t|-|h|), & |t-\Delta^2/t| < |h| \leq |t|, \\
|h|-|t|, & |h| > |t| \end{array} \right. \nonumber \\
\end{eqnarray}
To compute the gap in the presence of fluctuations, we numerically search for poles of the full Green's 
functions in the complex plane of $i\omega$.  The Matsubara formalism expressions we have
obtained for general inverse temperature $\beta$ can be analytically continued away from the
imaginary axis of $i\omega$ when we formally take the limit of zero temperature, $\beta \rightarrow \infty$.
We have developed a suite of numerical procedures to obtain the Green's functions
in this limit.
Poles of the full Green's function are then found by requiring explicitly that the denominator
(\ref{Dysondenom}) vanishes.  The pole with the lowest real value of $i\omega$ then determines
the position of the one-particle excitation gap.  Explicitly, we have written a numerical algorithm finding
roots of 
\begin{eqnarray}
a(k, \omega_r) a(k,-\omega_r) - b(k,\omega_r) b(k,-\omega_r) = 0, 
\end{eqnarray}
where
\begin{eqnarray}
&&a(\omega_r) = \Sigma_N(-\omega_r) + \mathcal{G}_0(k,\omega_r)/{\mbox{Det} \boldsymbol{\mathcal{G}}_0
(k, \omega_r)}, \nonumber \\
&&b(\omega_r) = \Sigma_S(\omega_r) + \mathcal{F}_0(k,\omega_r)/{\mbox{Det} \boldsymbol{\mathcal{G}}_0
(k, \omega_r)}.
\end{eqnarray}
An effective dispersion relation $\omega_r(k)$ is thus obtained, whose minimum can be found by numerically
scanning along momentum values.  The gap is then given by ($\mbox{Min}_k$ here meaning the minimization of the function with
respect to $k$) 
\begin{eqnarray}
\omega_{gap} = \left\{ \begin{array}{cc}
\mbox{Min}_k ~~2\omega_r, & |h| < |t|, \\
\mbox{Min}_k ~~~\omega_r, & |h| > |t|.
\end{array} \right.
\end{eqnarray}
Upon inspection of the results of these computations, some pleasant features emerge.  
The pure mean-field theory predicted a purely convex gap as a function of field below the
transition; in the presence of fluctuations, this prediction gets greatly modified
for fields below the critical field.  For example, in the case of figure (\ref{Gap2}), the
gap now reaches a maximal value somewhat below the transition, and decreases for decreasing
field, therefore reproducing basic features expected of an exact solution of the problem.
The zero-field gap (discussed in the next section) also agrees very well with the exact value:
in figure (\ref{Gap1}), the mean-field value at zero field was 22 percent away from the exact one,
whereas fluctuations bring this discrepancy right down by two orders of magnitude to 0.2 percent.
In figure (\ref{Gap2}), the same comparisons yield 98 percent versus 11 percent, again an increase
in precision by about an order of magnitude.

\subsection{A note on the numerics}
\begin{figure}
\includegraphics[width=8cm]{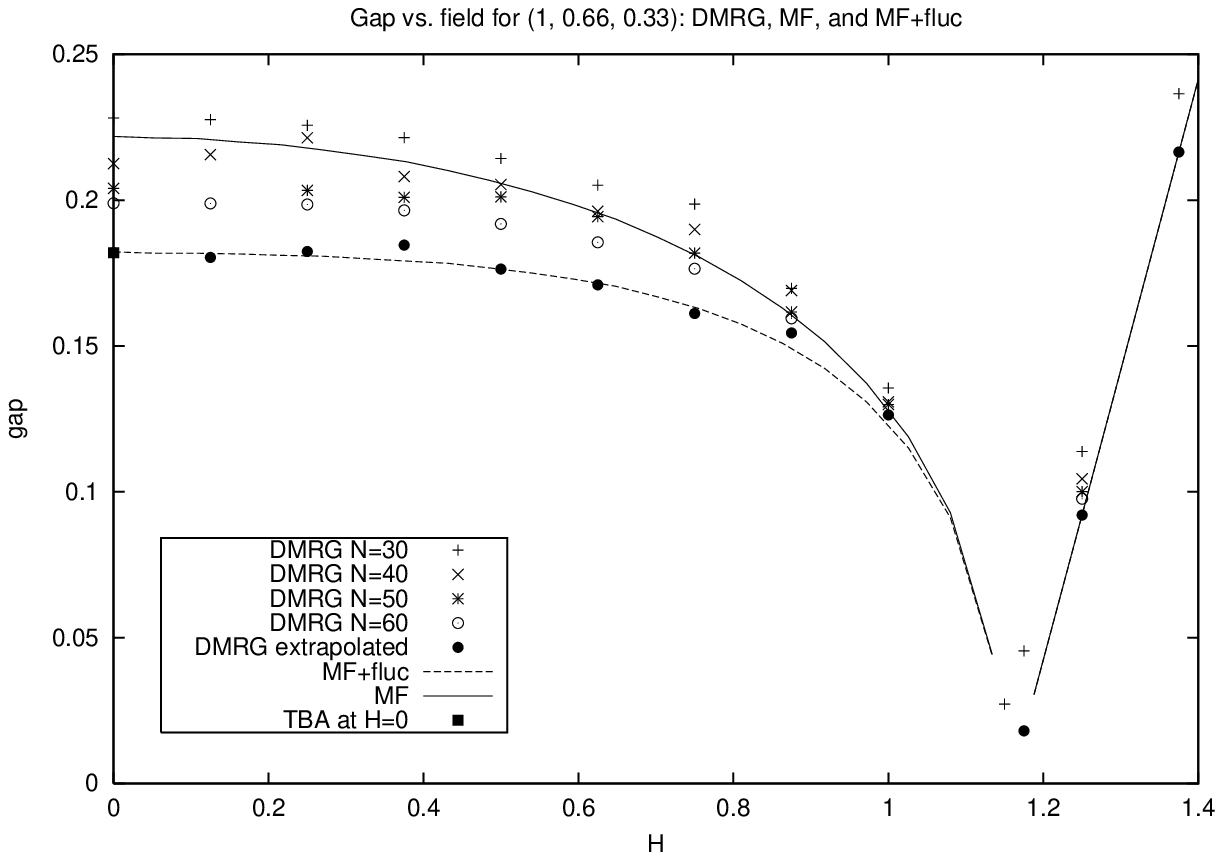}
\caption{Gap as a function of field, computed in mean-field and with fluctuations.  The dot at
zero field represents the Bethe Ansatz exact value.}
\label{Gap1}
\end{figure}

\begin{figure}
\includegraphics[width=8cm]{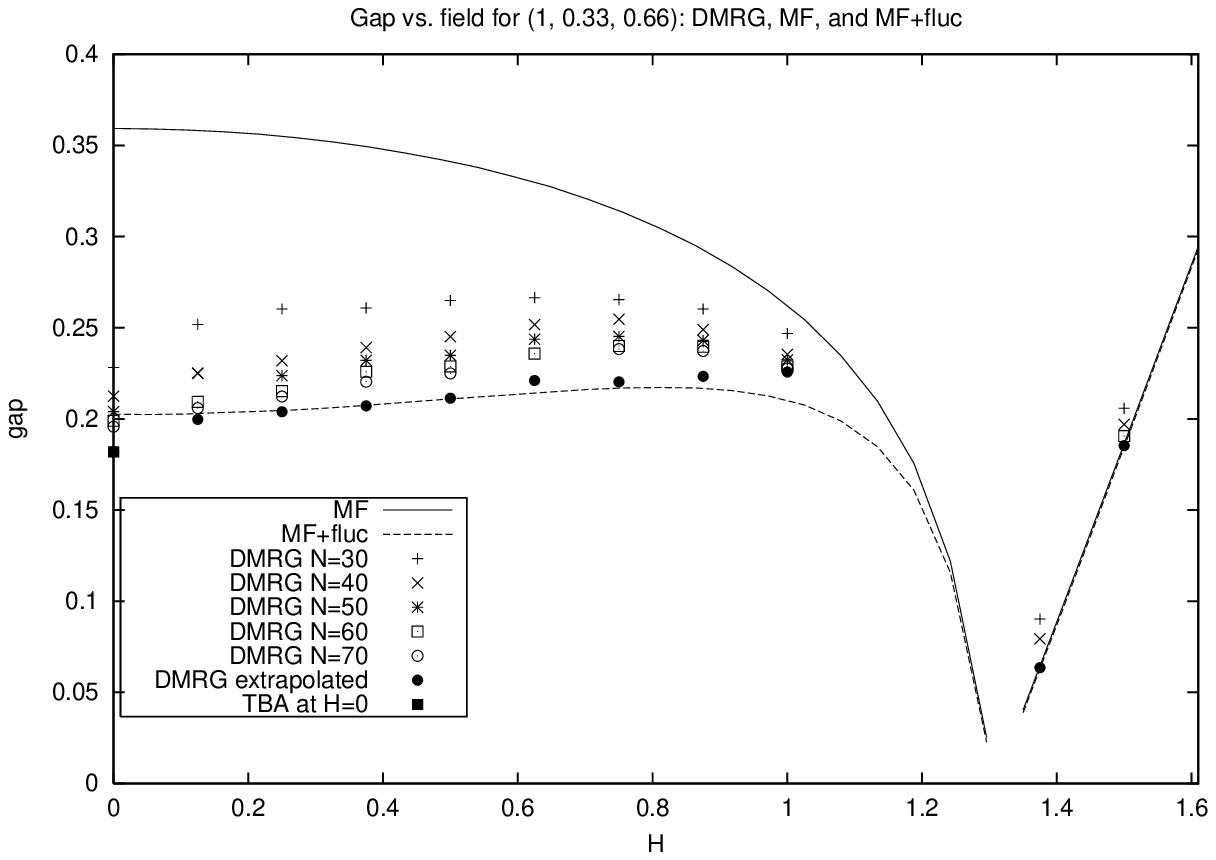}
\caption{Gap as a function of field, computed in mean-field and with fluctuations.  The dot at
zero field represents the Bethe Ansatz exact value.}
\label{Gap2}
\end{figure}

\begin{figure}
\includegraphics[width=8cm]{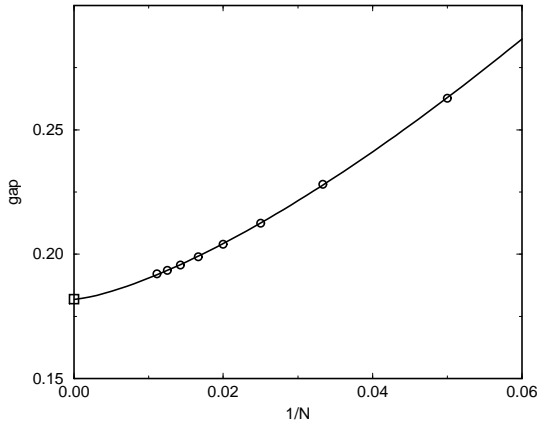}
\caption{Example of numerical data for the gap used in the extrapolation to $N \rightarrow \infty$, 
here for the zero-field XYZ magnet with exchange parameters $(1, 0.66, 0.33)$.  The circles represent DMRG data for
finite $N$, while the square is the exact zero-field value from the Bethe Ansatz.}
\label{Gap_extrap}
\end{figure}

We use the standard DMRG method with periodic boundary conditions
 \cite{White} as an independent numerical tool 
 to study the magnetizations and  excitation gaps of the Hamiltonian 
 Eq(1). We calculate the energies of the 
 three lowest states of the systems,
 for system sizes up to N=100 spins.
 As expected, for $H<H_c$ the gap between the lowest 
 and the first excited state vanishes rapidly with growing 
 system size reflecting the fact that the ground state is twofold 
 degenerate in the thermodynamic limit.
 In Figs (\ref{Gap1}, \ref{Gap2}) we do not show these vanishing gaps
 but only display the energy diffrence between the second 
 excited state and the lowest state.
 For a better comparison with the field theoretical 
 results we extrapolate our results 
 using systems of size $N=30...100$ spins. 
 As can be seen from Fig(\ref{Gap_extrap}) our data for the gap
 extrapolate well to the Bethe value, 
 though the effects of the finite system size is still 
 substantial even for 100 spins.
 The extrapolation procedure for non vanishing magnetic field
 $H<H_c$ comes with an uncertainty which we estimate to be 
 5 percent, this error is mostly due to the well known problem 
 of unsystematic finite size effects in nonzero magnetic-field.

 For $H>H_c$ the lowest gap is 
 the energy difference between the lowest and the first excited state.
 Finite size effects in this region are negligible.

\section{Bethe Ansatz for zero field}
The Hamiltonian of the zero-field, fully anisotropic $\mbox{XYZ}$ Heisenberg spin chain 
is related via trace identities to the
transfer matrix of the eight-vertex model, solved exactly by Baxter \cite{BaxterAP70}.
Let us recall some aspects of the exact solution of the model, to which our 
perturbative results at zero field can be compared.  For the zero-field anisotropic chain
\begin{eqnarray}
H \!=\! \sum_{j = 1}^N \left[ J_x S_j^x S_{j+1}^x \!+\! J_y S_j^y S_{j+1}^y 
\!+\! J_z S_j^z S_{j+1}^z \right],
\end{eqnarray}
we consider the case $J_z \geq J_y \geq |J_x| \geq 0$, which covers all parameter space for zero
field.  It is straightforward to map this restricted case to any of the cases we have
considered above.  

Excitations over the ground-state are formed by
spinons whose single-particle energy gap \cite{TakahashiBOOK} is given by
\begin{eqnarray}
\Delta_{sp} = J_z \frac{\mbox{sn}_l 2\zeta K(u') u'}{2\zeta}.
\end{eqnarray}
Our notations are such that
$\mbox{sn}$ is Jacobi's sinus amplitudinus function of modulus $k$, where
\begin{eqnarray}
k = \frac{1-l}{1+l}, \hspace{1cm} l = \sqrt{\frac{J_z^2 - J_y^2}{J_z^2 - J_x^2}},
\end{eqnarray}
and $\mbox{sn}_l$ is the function of modulus $l$.
The parameter $\zeta$ is given by
\begin{eqnarray}
\zeta = K(l) + \frac{2i \eta}{1+l}, 
\end{eqnarray}
where $K(l)$ is the complete elliptic integral of modulus $l$, and 
$\eta$ is determined from the anisotropy parameters of the model according to
\begin{eqnarray}
\mbox{sn}^2 2\eta = - \left( \frac{\sqrt{J_z^2 - J_x^2} + \sqrt{J_z^2 - J_y^2}}{J_y - J_x}
\right)^2.
\end{eqnarray}
Finally, the argument $u$ and its complementary $u' = \sqrt{1-u^2}$ are determined
from 
\begin{eqnarray}
\frac{K(u')}{K(u)} = \frac{\zeta}{K(l')}.
\end{eqnarray}

\begin{figure}
\includegraphics[width=8cm,height=7cm]{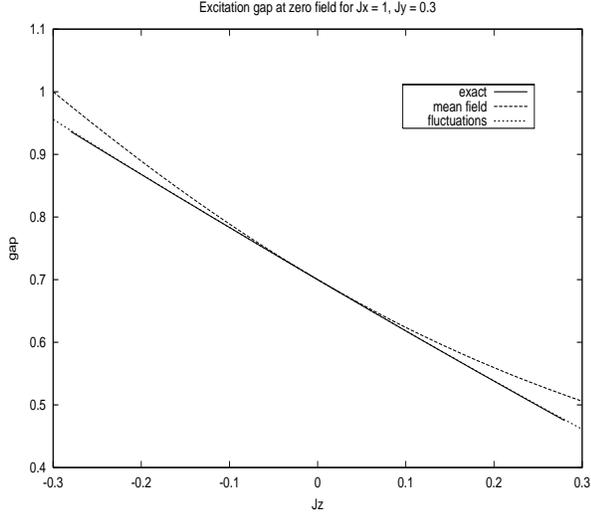}
\caption{Excitation gap as a function of the exchange parameter $J_z$, for strong
$x,y$ anisotropy.  The straight mean-field result lies up to 5 percent away from
the exact solution, whereas the fluctuations-corrected result precisely overlaps it.}
\label{Gapzero1}
\end{figure}

\begin{figure}
\includegraphics[width=8cm,height=7cm]{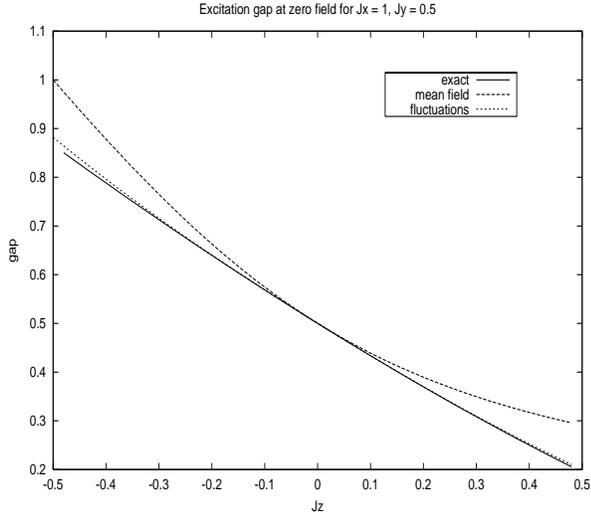}
\caption{Excitation gap as a function of the exchange parameter $J_z$, for weaker
$x,y$ anisotropy.  Once again, the fluctuations-corrected result greatly improves 
the mean-field one.}
\label{Gapzero2}
\end{figure}

\begin{figure}
\includegraphics[width=8cm,height=7cm]{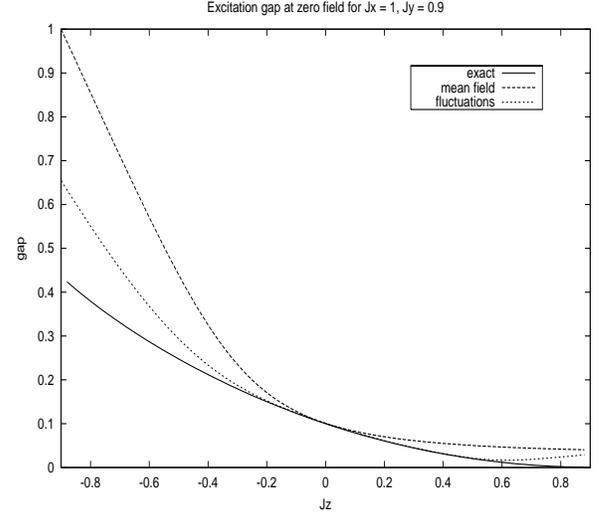}
\caption{Excitation gap as a function of the exchange parameter $J_z$, for weak
$x,y$ anisotropy.  The fluctuations-corrected result greatly improves the mean-field one, 
but the agreement is less spectacular than in the above two cases, due to the relative smallness
of the zero-field gap for weak anisotropy.}
\label{Gapzero3}
\end{figure}

Since we are by definition below the critical field, spinons are always created in pairs, 
and therefore the physical excitation gap at zero field is given by $2 \Delta_{sp}$.  
We can plot the exact value for the gap from the Bethe Ansatz against the mean-field
prediction, and the fluctuations-corrected prediction.  We present three graphs of
the gap as a function of the exchange parameter $J_z$, for three different choices
of $J_x$ and $J_y$.  In figure (\ref{Gapzero1}), we lie rather deep in the anisotropic limit,
where the zero-field gap is rather large throughout the parameter range.  
The agreement between mean-field and the exact result is acceptable (the
variation in the ferromagnetic end (negative $J_z$) being 4.9 percent and that at the antiferromagnetic
end being 8.6 percent), but
incorporating fluctuations makes the agreement dramatically better (down to 0.2 percent and 0.3 percent,
respectively, so comfortably over an order of magnitude increase in precision).
In figure (\ref{Gapzero2}), the $x,y$ anisotropy is reduced,
and therefore also the gap.  This leads to greater disagreement, although the exact 
solution and fluctuations-corrected results are still extremely close.  Here, the mean-field
strays away from the exact solution by 15 percent and 43 percent for the ferromagnetic and antiferromagnetic
ends, respectively, whereas the fluctuations results differ from the exact solution by only 1.8 percent
and 2.2 percent respectively.  Finally, 
figure (\ref{Gapzero3}) put us closer to the (gapless) isotropic case, and as expected
shows the greatest deviations.  For the purely antiferromagnetic case $J_z > 0$, however,
the fluctuations-corrected result reproduces the exact solution to great precision up to $J_z 
\approx 0.6$.

\begin{figure}
\includegraphics[width=8cm,height=7cm]{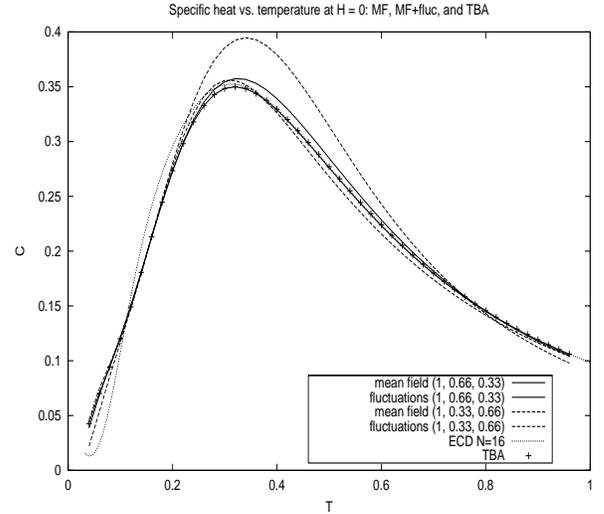}
\caption{Specific heat for an $XYZ$ magnet in zero field, computed from the Bethe Ansatz and
compared to the straight mean-field and fluctuations results obtained in two different
ways (interchanging the $Y$ and $Z$ anisotropies, {\it i.e.} our perturbation parameter).  The
fluctuations result for $J_z = 0.33$ is indistinguishable from the exact result on the 
resulting plot.}
\label{tba-heat}
\end{figure}

The Bethe Ansatz allows also the computation of thermodynamic quantities in the model.
The free energy at temperature $T$ can be written
\begin{eqnarray}
f = - T \ln \Lambda_0, 
\end{eqnarray}
where $\Lambda_0$ is the largest eigenvalue of the transfer matrix of the model.  Its
value is determined as
\begin{eqnarray}
\Lambda_0 = 2 \prod_{j = 1}^{\infty} \left( \frac{J_z}{4 \pi T (j - 1/2)} \right)^2
(1 + p_j^2)
\end{eqnarray}
where the $p_j$ parameters satisfy the coupled equations
\begin{eqnarray}
\frac{J_z}{2T} G (p_l) \!&\!-\!&\! \sum_{j = 1}^{\infty} \left[ \phi (p_l, p_j) + \phi (p_l, - p_j) \right] 
= 2\pi (l - 1/2), \nonumber \\
G(p) \!&\!=\!&\! p \frac{\sqrt{(p^2 + D_x^2)(p^2 + D_y^2)}}{(p^2 + 1)}, \nonumber \\
\phi(p, q) \!&\!=\!&\! 2 \arctan \left( \frac{p \Delta_x \Delta_y + q (1 + p^2)}
{\sqrt{(p^2 + D_x^2)(p^2 + D_y^2)}} \right),
\end{eqnarray}
in which the anisotropy parameters appear as $\Delta_x = J_x/J_z, ~\Delta_y = J_y/J_z$, 
and $D_x = \sqrt{1 - \Delta_x^2}, ~D_y = \sqrt{1 - \Delta_y^2}$.  These equations can
be solved numerically, which permits the computation of the temperature-
dependent specific heat of the model at zero field.  A comparison of our perturbative
results with the exact TBA result is given in figure (\ref{tba-heat}).

\begin{figure}
\includegraphics[width=8cm,height=7cm]{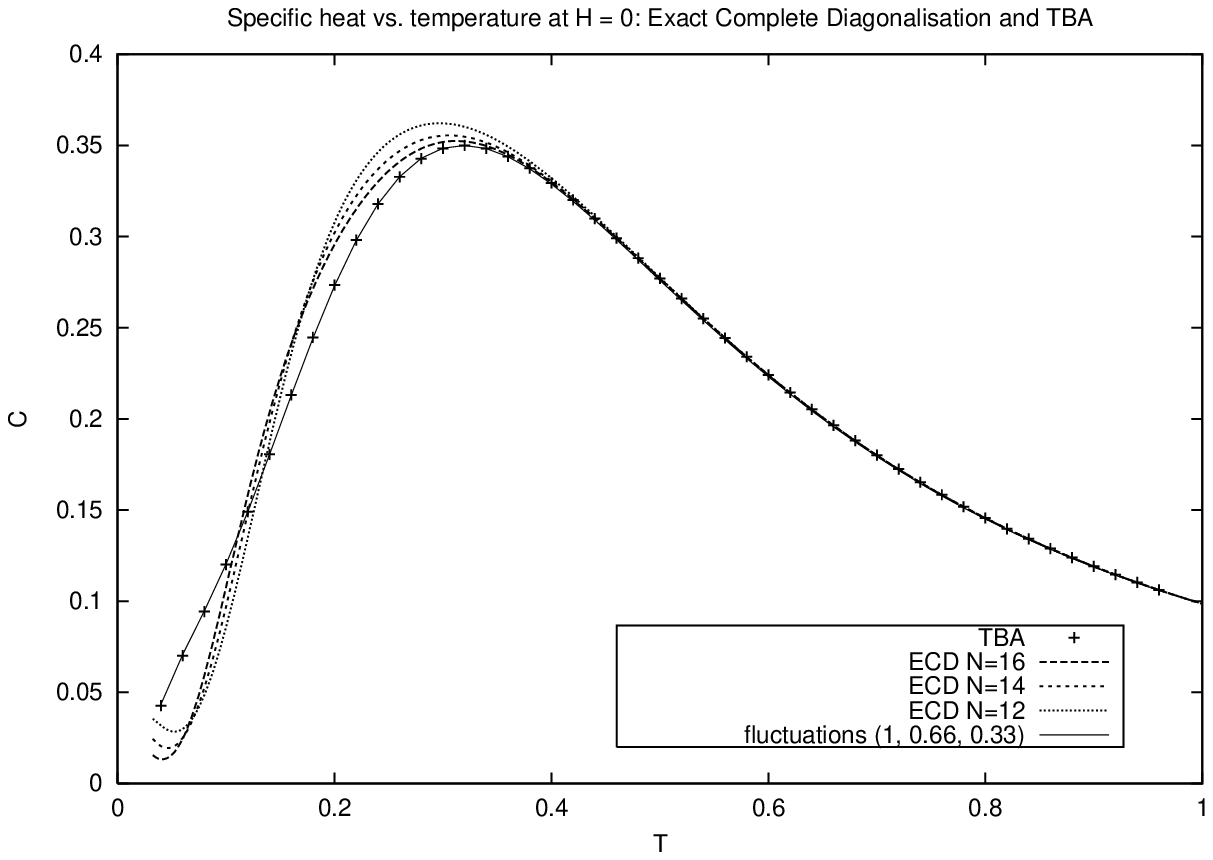}
\caption{Specific heat for the same case as in figure (\ref{tba-heat}), computed from
exact complete diagonalization for $12, 14$ and $16$ sites.  The fluctuations and exact TBA 
curves are included for comparison.}
\label{tba-heat2}
\end{figure}

\section{Conclusion}
While the perturbative scheme that we have developed and implemented here does not constitute
a definitive treatment of anisotropic Heisenberg chains in a field, which can only be achieved
through a nonperturbative computation of Bethe Ansatz-type, it does provide an efficient and
credible approach to the computation of approximate numerical values for physical quantities 
in general.  The comparisons with DMRG and (for zero field) with the Bethe Ansatz that we have
provided are a good guide to establishing the relative numerical precision to be expected from
our theory, which ranges from very good for the ``worst-case'' scenario of a gapless $\mbox{XXZ}$
chain, to essentially indistinguishable from the exact solution (errors of order 0.2 percent) 
for spin chains with strong structural $\mbox{XYZ}$ anisotropy.  In broad terms, an increase
in precision of at least an order of magnitude is achieved by incorporating the fluctuations 
as we have done here, which makes our approach good enough for many practical purposes.  Here,
we have concentrated on equilibrium thermodynamic quantities like magnetization, and basic
dynamical-related ones like the excitation gap.  It would be of great interest to extend
our treatment to purely dynamical ones like the spin structure factor, which we will do in a
forthcoming publication. 
\\
\acknowledgments
It is a pleasure to thank R. Coldea, C. Mukherjee, D. A. Tennant and F. H. L. Essler for stimulating discussions. 
J.-S. C. acknowledges support from the Dutch FOM foundation.

\end{document}